\documentclass[aps,prl,twocolumn,showpacs,amsmath,amssymb,superscriptaddress]{revtex4}
\usepackage{graphicx}
\usepackage{amssymb}
\usepackage{natbib}
\usepackage{color}

\begin{document}
\title{Double neutron spin resonances and gap anisotropy in underdoped superconducting 
NaFe$_{0.985}$Co$_{0.015}$As}

\author{Chenglin Zhang}
\affiliation{Department of Physics and Astronomy, Rice University, Houston, Texas 77005, USA}

\affiliation{Department of Physics and
Astronomy, The University of Tennessee, Knoxville, Tennessee
37996-1200, USA}

\author{Rong Yu}
\affiliation{Department of Physics and Astronomy, Rice University, Houston, Texas 77005, USA}

\author{Yixi Su}
\affiliation{J\"{u}lich Centre for Neutron Science, Forschungszentrum J\"{u}lich GmbH, Outstation at FRM II, Lichtenbergstrasse 1, D-85747 Garching, Germany
}
\author{Yu Song}
\affiliation{Department of Physics and Astronomy, Rice University, Houston, Texas 77005, USA}
\affiliation{Department of Physics and
Astronomy, The University of Tennessee, Knoxville, Tennessee
37996-1200, USA}

\author{Miaoyin Wang}
\affiliation{Department of Physics and Astronomy, The University of
Tennessee, Knoxville, Tennessee 37996-1200, USA}

\author{Guotai Tan}
\affiliation{Department of
Physics and Astronomy, The University of Tennessee, Knoxville,
Tennessee 37996-1200, USA}
\affiliation{Physics department, Beijing
Normal University, Beijing 100875, China}
\author{Takeshi Egami}
 \affiliation{Department of Physics and Astronomy, The University of Tennessee,
Knoxville, Tennessee 37996-1200, USA}
\affiliation{Department of Materials Science and Engineering, The University of Tennessee,
Knoxville, Tennessee 37996-1200, USA}
\affiliation{Oak Ridge National Laboratory,
Oak Ridge, Tennessee 37831, USA}
\author{J. A. Fernandez-Baca}
\affiliation{Oak Ridge National Laboratory,
Oak Ridge, Tennessee 37831, USA}
\affiliation{Department of Physics and
Astronomy, The University of Tennessee, Knoxville, Tennessee
37996-1200, USA}
\author{Enrico Faulhaber}
\affiliation{Gemeinsame Forschergruppe HZB - TU Dresden,
Helmholtz-Zentrum Berlin f\"{u}r Materialien und Energie, D-14109 Berlin,
Germany} \affiliation{Forschungsneutronenquelle Heinz Maier-Leibnitz
(FRM-II), TU M\"{u}nchen, D-85747 Garching, Germany}

\author{Qimiao Si}
\affiliation{Department of Physics and Astronomy, Rice University, Houston, Texas 77005, USA}

\author{Pengcheng Dai} 
\email{pdai@rice.edu}
\affiliation{Department of Physics and Astronomy, Rice University, Houston, Texas 77005, USA}
\affiliation{Department of Physics and Astronomy, The University of Tennessee,
Knoxville, Tennessee 37996-1200, USA}

\begin{abstract}
We use inelastic neutron scattering to show that superconductivity in electron-underdoped 
NaFe$_{0.985}$Co$_{0.015}$As induces a dispersive sharp resonance near $\textit{E}_{r1} = 3.25$ meV and a broad
dispersionless mode at $\textit{E}_{r2} = 6$ meV.
However, similar measurements on overdoped superconducting NaFe$_{0.955}$Co$_{0.045}$As 
find only a single sharp resonance at $E_r=7$ meV.  
We connect these results with the observations of angle-resolved photoemission spectroscopy
that the superconducting gaps in the electron Fermi pockets are anisotropic in the
underdoped material but become isotropic in the overdoped case.
Our analysis indicates that both the double neutron spin resonances and gap anisotropy originate
from the orbital dependence of the superconducting pairing in the iron pnictides. 
Our discovery also shows the importance of the inelastic neutron scattering in detecting 
the multiorbital superconducting gap structures of iron pnictides. 
\end{abstract}

\pacs{74.25.Ha, 74.70.-b, 78.70.Nx}

\maketitle


High-transition temperature (high-$T_c$) superconductivity in copper oxides and iron pnictdies 
can be derived from electron or hole doping to their antiferromagnetic (AF) parent compounds \cite{jmtranquada,dai}.  
Since magnetism may be a common thread for the electron pairing in high-$T_c$ superconductors \cite{scalapino}, 
it is important to determine how magnetic excitations can probe the superconducting (SC) 
electron pairing interactions.  For single band copper oxide superconductors, the neutron spin resonance, 
a sharp collective magnetic excitation 
at the AF ordering wave vector below $T_c$, 
has been the subject of twenty years' study and 
provided strong evidence for the sign changing nature 
 of the $d$-wave superconducting gap in these materials \cite{Eschrig}.  
 In the case of multiband iron pnictide superconductors \cite{kamihara,cwchu}, band structure calculations 
 indicate that the Fermi surfaces consist of hole pockets near the zone center and electron pockets 
 near the zone corner \cite{Singh,kuroki,Mazin,hirschfeld,chubukov}.  
 Although the sign change of the quasipartice excitations (nesting) between the hole and electron pockets 
 also necessitates a resonance at an energy below the sum of the electron 
 and hole SC gap energies \cite{Korshunov,Maier}, 
 the multiple 3$d$ Fe orbital nature of the iron pnictides \cite{Yi09,Yi13} means that the SC gaps 
 can be anisotropic on different Fermi surfaces \cite{Maier09,Goswami}.  
Therefore, if the resonance is a direct probe of the quasiparticle excitations between the
hole and electron Fermi pockets, it should be sensitive to the SC gap 
energy anisotropy.  In spite of intensive inelastic neutron 
scattering (INS) work on hole \cite{christianson,chenglinzhang,castellan} 
and electron-doped \cite{lumsden,schi09,dsinosov09,steffens} BaFe$_2$As$_2$ family of iron pnictides, 
only a broad resonance consistent with 
the sign change of the SC pairing has been observed.

For the NaFe$_{1-x}$Co$_x$As family of iron pnictides [Fig. 1(a)] \cite{slli09,Parker}, 
the London penetration depth measurements suggest that the SC gap is highly
anisotropic even at optimal doping \cite{cho}.  Moreover,
 angle-resolved photoemission 
(ARPES) experiments indicate the presence of large SC gap anisotropy 
in the electron Fermi pockets of the underdoped 
regime near $x=0.0175$, which is absent in the hole Fermi pockets; the gap anisotropy
disappears upon increasing $x$ to 0.045 [Figs. 1(c),1(d)] \cite{qqge,Liu_arpes,thirupathaiah}.
A likely origin \cite{RYu_theory}
of this gap anisotropy is the angular variation of the relative orbital weight among
the $d_{xy}$ and  the degenerate $d_{xz/yz}$ orbitals along the electron Fermi pockets,
which is absent 
along the hole Fermi pockets. As such, this material
offers the opportunity to study the role of orbital dependence in SC pairing via INS.

In this Letter, we present INS study of spin excitations in
underdoped SC NaFe$_{0.985}$Co$_{0.015}$As coexisting with static AF order ($T_c=15$ K, $T_N=30$ K) and its comparison
with overdoped SC NaFe$_{0.955}$Co$_{0.045}$As ($T_c=20$ K) 
[Fig. 1(a)] \cite{clzhang13}.
Our INS experiments reveal that superconductivity induces two distinct neutron spin resonances
at the commensurate AF wave vector ${\bf Q} = (0.5, 0.5,L)$ 
in NaFe$_{0.985}$Co$_{0.015}$As [Figs. 2(a-c)];
this is an entirely new behavior which has never been observed 
in either the iron-based or copper-based
superconductors.
While the first resonance occuring at $E_{r1}=3.25$ meV is sharp in energy and becomes
dispersive along the $c$-axis,
there is also a broad dispersionless resonance at $E_{r2}=6$ meV [Figs. 2(e-g)].
For electron-overdoped SC NaFe$_{0.955}$Co$_{0.045}$As, the double resonances
changes back to a single resonance [Fig. 1(f)] \cite{clzhang13}. 
Our analysis
indicates 
that both the SC gap anisotropy and the double resonances arise from the orbital 
dependent pairing strength,
and reveals the important role that INS can play in probing of the 
multiorbital structure of superconductivity in the iron-based superconductors.

\begin{figure}[t] \includegraphics[scale=.4]{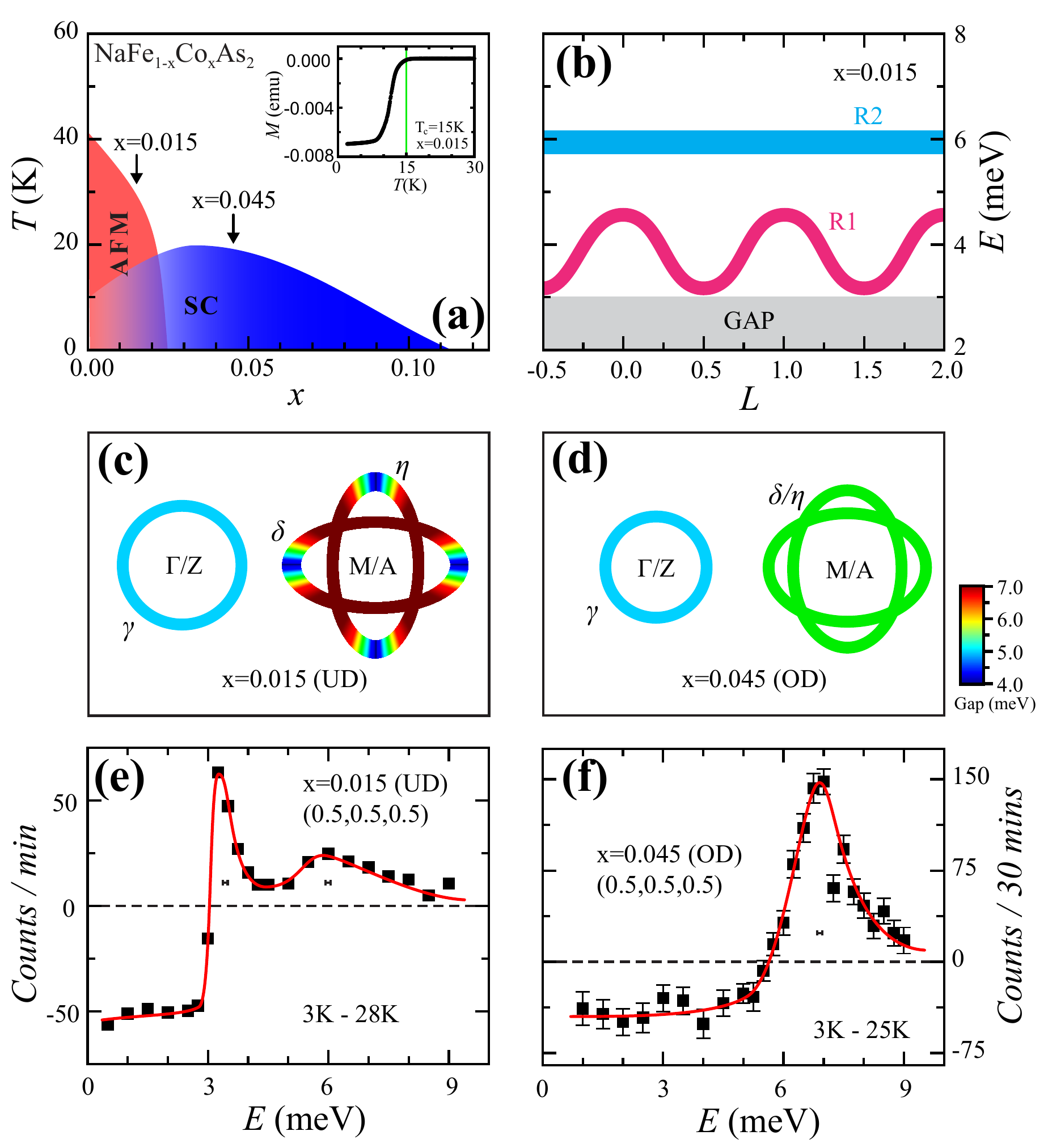}
\caption{
(a) The electronic phase diagram of NaFe$_{1-x}$Co$_x$As, where the arrows indicate the Co-doping levels of our samples.  The temperature dependence of the bulk susceptibility in the inset shows
$T_c=15$ K.
(b) The schematics of the $c$-axis dispersion of the double resonances.
(c,d) The schematics of Fermi surfaces and SC gaps in underdoped and overdoped
samples near $\Gamma$ and $M$ points \cite{qqge}.
(e) Double resonances obtained by taking temperature difference plots (4 K$-$28 K)
of constant-$Q$ scans at $(0.5,0.5,0.5)$
 in NaFe$_{0.985}$Co$_{0.015}$As. (f) Similar data in NaFe$_{0.955}$Co$_{0.045}$As showing only
 a single resonance. The horizontal bars in (e) and (f) indicate instrumental energy resolution.}
\end{figure}

We prepared $\sim$5 g single crystals of NaFe$_{0.985}$Co$_{0.015}$As by self-flux method \cite{clzhang13}.  Susceptibility [inset in Fig. 1(a)],
heat capacity \cite{gttan}, and nuclear magnetic resonance \cite{swoh}
measurements showed that the sample is a homogeneous bulk superconductor ($T_c=15$ K) microscopically
coexisting with static AF order.
Our neutron scattering experiments were carried out on the thermal (HB-3) and cold (PANDA)
triple-axis spectrometers at High Flux Isotope Reactor, Oak Ridge
National Laboratory and the FRM-II, TU M\"{u}chen, Germany \cite{schi09}, respectively.
At HB-3, we fixed final neutron energies at $E_f=14.7$ meV with Pyrolytic graphite (PG) monochromator 
and analyzer.
At PANDA, We used focusing PG monochromator and analyzer with a fixed final neutron energy of $E_f=5$ meV.
The wave vector ${\bf Q}$ at ($q_x$,$q_y$,$q_z$) in \AA$^{-1}$ is
defined as (\textit{H},\textit{K},\textit{L}) = ($q_xa/2\pi$,$q_ya/2\pi$,$q_zc/2\pi$) reciprocal lattice unit (rlu) 
using the
tetragonal unit cell (space group $P4/nmm$, $a \approx 3.952$ \AA\ and $c = 6.980$ \AA\ at 3 K).  In this notation,
the AF Bragg peaks occur at the $(0.5,0.5,L)$ positions with $L=0.5,1.5,\cdots$ \cite{slli09}.
The samples are
coaligned in the $[H,H,L]$ scattering zones with a mosaic less than 2$^\circ$.  
Figure 4(a) shows the temperature dependence
of the elastic scattering at ${\bf Q}_{AF}=(0.5,0.5,0.5)$,
which reveals a clear reduction at the onset of $T_c$ and dramatic increase 
below $T_N=30$ K [Fig. 4(a)]. These results suggest that NaFe$_{0.985}$Co$_{0.015}$As is a homogeneous electron underdoped superconductor
similar to underdoped SC BaFe$_{2-x}T_x$As$_2$ ($T=$Co, Ni) [Fig. 1(a)] \cite{pratt09,Christianson09}.
From earlier ARPES measurements \cite{qqge,Liu_arpes,thirupathaiah}, we know that the SC gaps 
in the electron and hole pockets are quite
isotropic for electron overdoped NaFe$_{0.935}$Co$_{0.045}$As [Fig. 1(d)], but the SC gap becomes 
highly anisotropic for NaFe$_{0.985}$Co$_{0.015}$As [Fig. 1(c)].

\begin{figure}[t] \includegraphics[scale=.33]{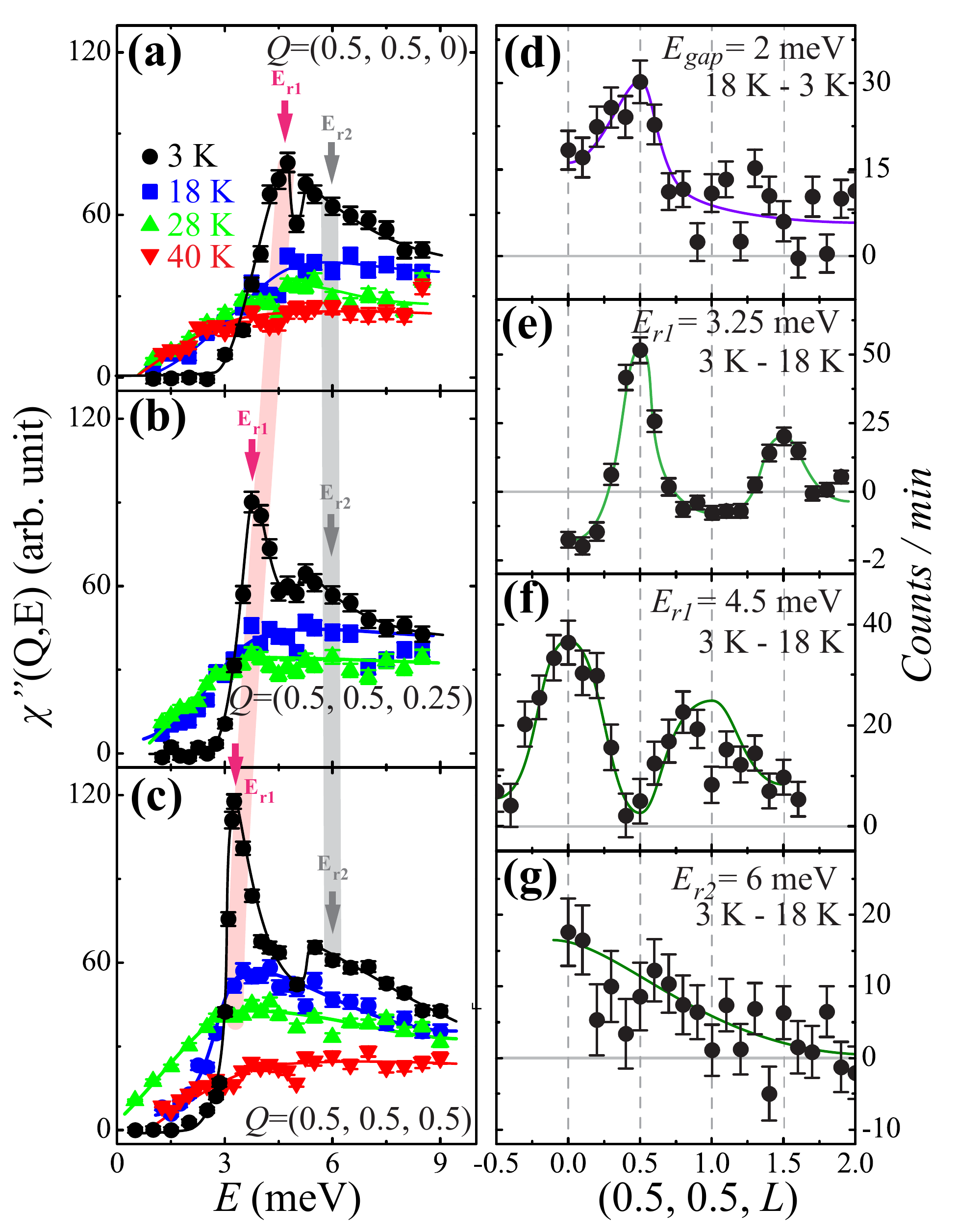}
\caption{
(a-c) $\chi^{\prime\prime}(Q,E)$ 
at $\textbf{Q}=(0.5,0.5,L)$ with \textit{L} = 0, 0.25 and 0.5, respectively, at 3, 18, 28 and 40 K.
(d-f) The difference of \textit{L}-modulations above and below $T_c$ at $E=2$, 3.25, 4.5 and 6 meV, respectively.
}
\end{figure}

\begin{figure}[t] \includegraphics[scale=.35]{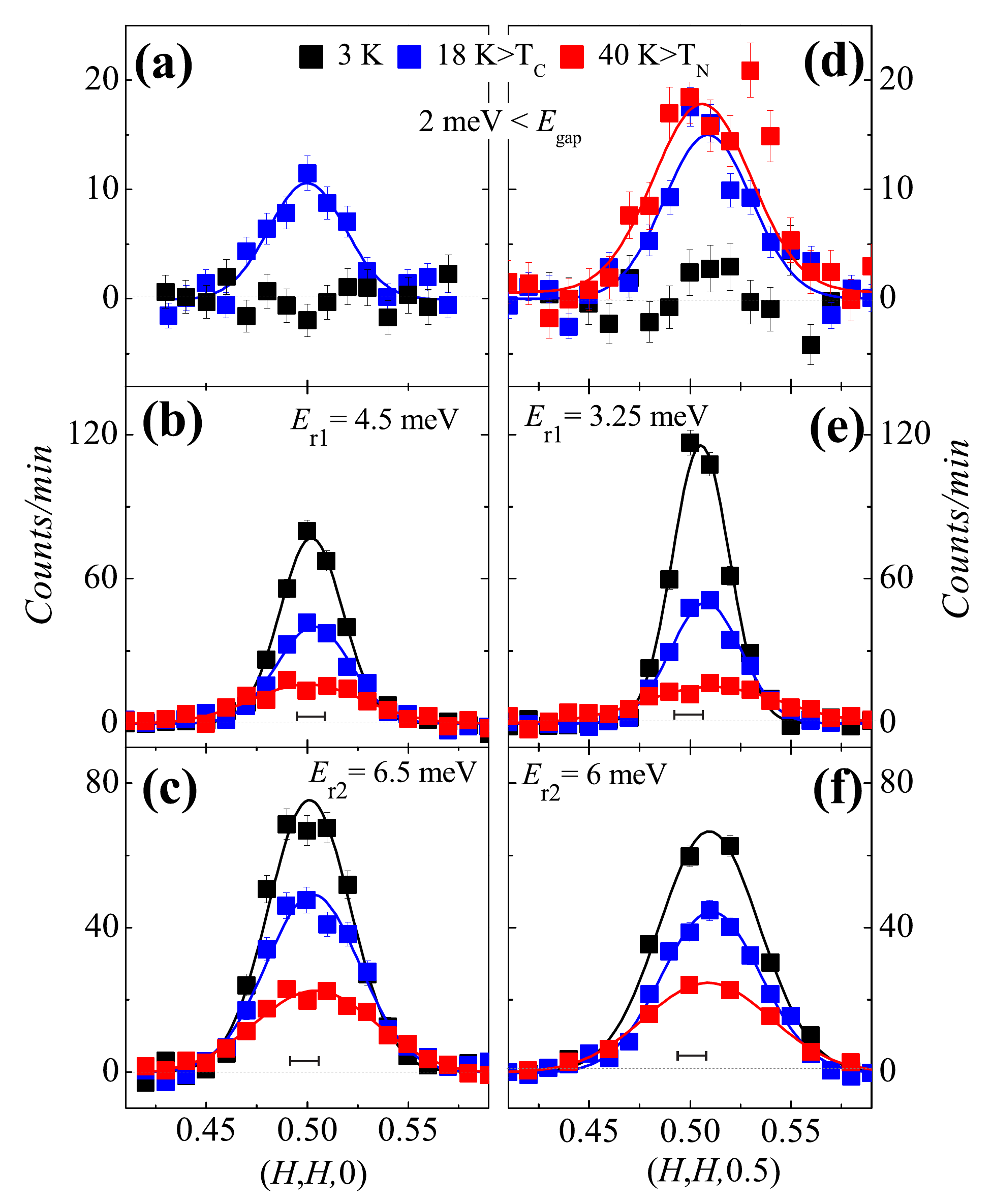}
\caption{
(a-c) $\bf Q$ scans along the $[\textit{H},\textit{H},0]$ direction at $E=2$ meV,  $E_{r1} = 4.5$ meV,
and $E_{r2} = 6.5$ meV, respectively, with $L=0$.
(d-f) $\bf Q$ scans along the $[\textit{H},\textit{H},0.5]$ direction at $E=2$ meV, $E_{r1}$ = 3.25 meV, 
and $E_{r2}$ = 6 meV, respectively, for SC NaFe$_{0.985}$Co$_{0.015}$As.
The horizontal bars indicate instrumental resolution.
The solid lines are fits to Gaussians.}
\end{figure}

In previous INS work on overdoped NaFe$_{0.935}$Co$_{0.045}$As ($T_c=18$ K), a dispersionless sharp 
resonance was found at $E_r=7$ meV below $T_c$ [Fig. 1(f)] \cite{clzhang13}.  To explore what happens in the underdoped 
regime where superconductivity coexists with
AF static order \cite{swoh}, we carried out constant-\textbf{Q} scans at wave vectors ${\bf Q}=(0.5,0.5,L)$ with $L = 0, 0.25$,  
and $0.5$ rlu
at $T<T_c$, $T_c<T<T_N$, and $T>T_N$ on NaFe$_{0.985}$Co$_{0.015}$As.
Figures 2(a)-2(c) show the $\chi^{\prime\prime}(Q,E)$ at $T=2, 18, 28, 40$ K, obtained by subtracting the 
background scattering of \textbf{Q}-scans in Fig. 3
and correcting for the Bose population factor using $\chi^{\prime\prime}(Q,E)=[1-\exp(-E/k_BT)]S(Q,E)$,
where $S(Q,E)$ is the magnetic scattering function.
At $T=40$ K ($T=T_N+10$ K), the paramagnetic scattering at all three wave vectors probed
are relaxational and can be fitted with $\chi^{\prime\prime}(Q,E)\propto E/(\Gamma^2+E^2)$
as shown in solid lines in Figs. 2(a)-2(c).  On cooling to $T=28$ K ($T=T_N-2$ K),
the overall lineshape of the scattering remain unchanged.  On further cooling to $T=18$ K ($T=T_c+3$ K), while
the scattering at wave vectors ${\bf Q}=(0.5,0.5,L)$ with $L = 0, 0.25$ still have Lorentzian lineshape (relaxational) 
[blue symbols in Figs. 2(a) and 2(b)],
a spin anisotropy gap of $\sim$1.5 meV opens at ${\bf Q}_{AF}=(0.5,0.5,0.5)$ 
[blue symbols in Fig. 2(c)].
Finally, upon entering into the SC state at $T=4$ K ($T=T_c-11$ K), 
we see that a sharp resonance and a broad resonance
develop at $E_{r1}=3.25$ and $E_{r2}=6$ meV, respectively, at ${\bf Q}_{AF}=(0.5,0.5,0.5)$ [Fig. 2(c)].  
In addition, the normal state spin gap of
$\sim$1.5 meV increases to $\sim$3 meV below $T_c$ [Fig. 2(c)].
The temperature difference plot between 4 K and 11 K shown in Fig. 1(e) confirms the presence
of superconductivity-induced double resonance.
 On changing wave vectors to
 ${\bf Q}=(0.5,0.5,0.25)$ and  ${\bf Q}=(0.5,0.5,0)$, we see that a clear increase in energy of the sharp 
 resonance while the broad mode remains at $E_{r2}=6$ meV [Figs. 2(b) and
 2(a)]. However, the low-temperature spin gaps are similar at all wave vectors.

To probe the $c$-axis modulations of the low-energy spin excitations and
superconductivity-induced effect, we carried out constant-energy scans
along the $[0.5,0.5,L]$ direction at different energies
above and below $T_c$.  Since there is a low-temperature
spin gap below $\sim$ 3 meV, the $L$-dependence of the
normal state magnetic scattering at $E=2$ meV
can be obtained by subtracting the data at $T=4$ K from those at
18 K.  The magnetic scattering at $E=2$ meV and 18 K shows a broad peak at
${\bf Q}_{AF}$ with $L=0.5$ rlu [Fig. 2(d)].
At the first resonance energy ($E_{r1}=3.25$ meV),
superconductivity induces well-defined peaks centered at 
 ${\bf Q}_{AF}=[0.5,0.5,L]$ with $L=0.5,1.5$ [Fig. 2(e)].
The energy of the first resonance moves to $E_{r1}=4.5$ meV at
$[0.5,0.5,L]$ with $L=0,1$, as illustrated in Fig. 2(f).
Figure 2(g) shows that the second resonance at $E_r=6$ meV is indeed dispersionless
with superconductivity-induced enhancement below $T_c$ decreases
monotonically with increasing
$L$, following the Fe magnetic form factor.

To confirm the low-temperature spin gap and determine
the wave vector dependence of the resonances, we carried out
constant-energy scans at diferent energies above and below $T_c$, and above $T_N$.
Figures 3(a-c) and 3(d-f) show $S(Q,E)$ along the $[\textit{H},\textit{H},0]$ and
$[\textit{H},\textit{H},0.5]$ directions, respectively.
At $E=2$ meV, a well-defined Gaussian peak in the normal state disappears below $T_c$,
confirming the presence of the low-temperature spin gap [Figs. 3(a) and 3(d)].
Comparing $E_{r1}=4.5$ meV with $L=0$ [Fig. 3(b)] and $E_{r1}=3.25$ meV with $L=0.5$
[Fig. 3(e)], we see that
the intensity gain of the resonances below $T_c$ is larger at $L=0.5$.
At the second resonance energy $E_{r2}=6$ meV [Figs. 3{e} and 3(f)],
superconductivity-induced intensity gain decreases with increasing $L$.
By Fourier transforming the fitted Guassian peaks, we find that
the spin-spin correlation lengths above $T_N$ are $\xi=33\pm 2$ \AA\ at $L=0,0.5$.
At 4 K and $L=0$, spin correlation lengths increase to $\xi=67\pm 2$ and $52\pm 2$ \AA\  at
$E_{r1}=4.5$ and $E_{r2}=6.5$ meV, respectively.
At 4 K and $L=0.5$, they are $75\pm 2$ and $42\pm 2$ \AA\ at 3.25 and 6 meV, respectively.

Figures 4(a-f) summarize the temperature dependence of the scattering at
different energies and wave vectors.  At the elastic AF Bragg position, we see
clear effect of $T_N$ and $T_c$ [Fig. 4(a)].
For $E=2$ meV, spin excitations show a kink at $T_N$ signaling the
static AF order, and decrease on cooling below $T_c$ [Figs. 4(b) and 4(c)].
From Figs. 4(d-f), we see that while the intensity at resonance energies
show kinks at $T_N$, they increase dramatically below $T_c$.
These results provide conclusive evidence of the presene of double resonance in underdoped
NaFe$_{0.985}$Co$_{0.015}$As.

\begin{figure}[t] \includegraphics[scale=0.38]{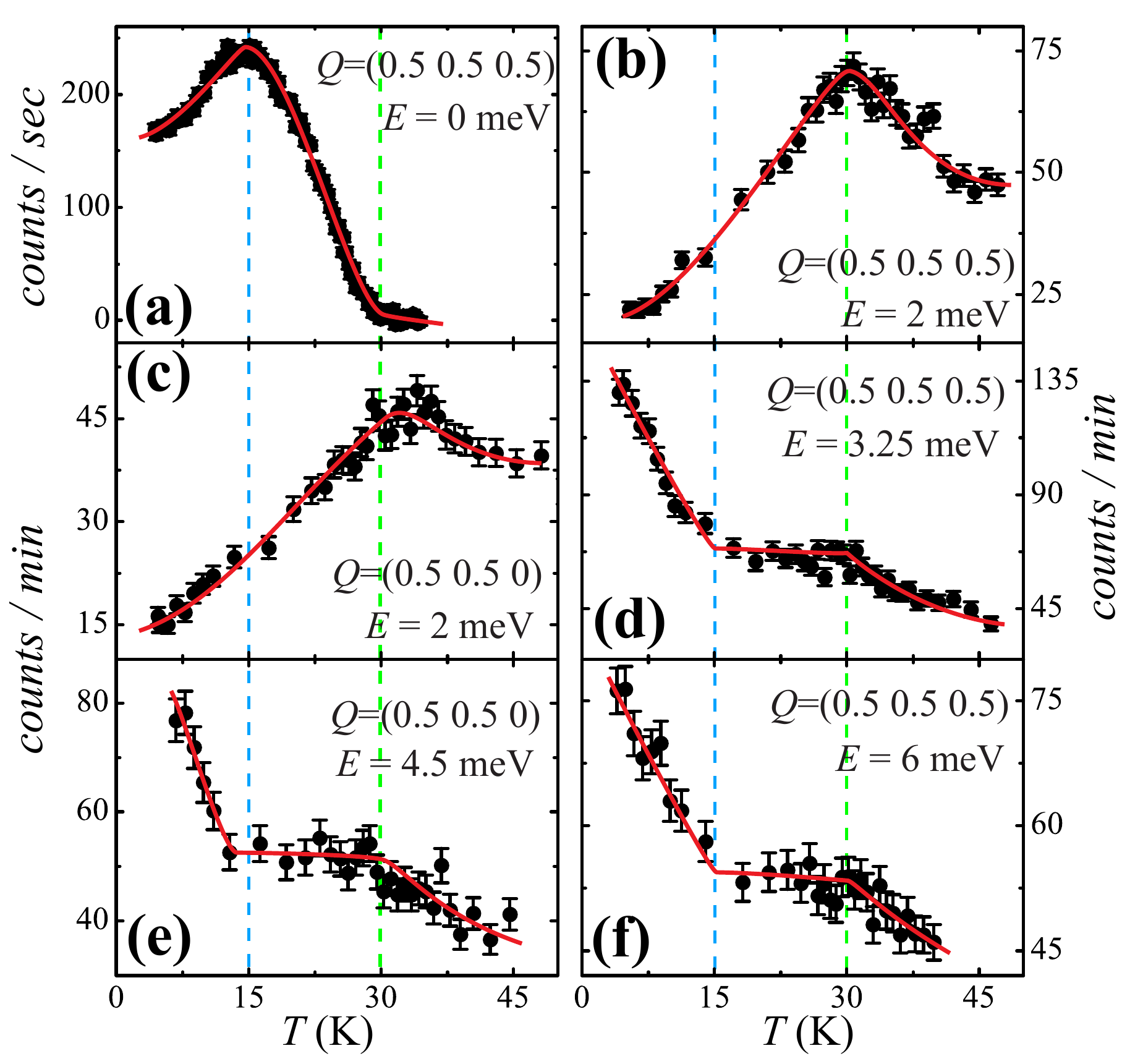}
\caption{
(a) The temperature dependence of
AF peak intensity at $\textbf{Q}=(0.5,0.5,0.5)$ with vertical dashed line indicating $T_c = 15$ K and $T_N = 30$ K.
(b) and (c) Temperature dependence of the scattering at $E=2$ meV at $\textbf{Q}=(0.5,0.5,0)$ and $(0.5,0.5,0.5)$, respectively.
Temperature dependence of the scattering at (d) $E_{r1}=3.25$ meV and
$(0.5,0.5,0.5)$, (e) $E_{r2}=4.5$ meV and
$(0.5,0.5,0)$, and (f) $E_{r2}=6$ meV and
$(0.5,0.5,0.5)$.
}
\end{figure}

In iron pnictides, the Fermi surface is composed of multiple orbitals. 
In electron doped NaFe$_{1-x}$Co$_x$As, the dominant orbital 
character of the electron pockets would be either $d_{xy}$ or $d_{xz/yz}$, depending on the direction 
in the Brillouin zone [Figs. 1(c) and 1(d)] \cite{qqge,Liu_arpes,thirupathaiah}. Recent theories and experiments 
find that the strength of electron correlations can be very different between the $d_{xy}$ and $d_{xz/yz}$ 
orbitals~\cite{Yu_multi11,YinKotliar11,Yu13b,Yi13}. This may induce orbital-selective SC pairing strengths, 
which naturally give anisotropic SC gaps along the electron pockets. The neutron resonance 
in the SC state is a bound state at energies just below the particle-hole excitation energy $E_r\leq\Delta_h + \Delta_e$ \cite{Eschrig}. 
If the anisotropic SC gap in the electron pocket is large, as in the underdoped NaFe$_{0.985}$Co$_{0.015}$As, 
there are two characteristic gaps $\Delta_{e1}\neq\Delta_{e2}$ (respectively associated with $d_{xy}$ 
and $d_{xz/yz}$ orbitals).
Two resonance peaks are 
expected as a result of this separation of energy scales. As the electron doping is increased to the overdoped regime, 
the orbital selectivity of the correlations is reduced~\cite{Yu13b}, which would give rise to a smaller SC gap anisotropy 
with $\Delta_{e1}\approx\Delta_{e2}$. Therefore, only one resonance peak would be resolved.

\begin{figure}[t] \includegraphics[scale=0.30]{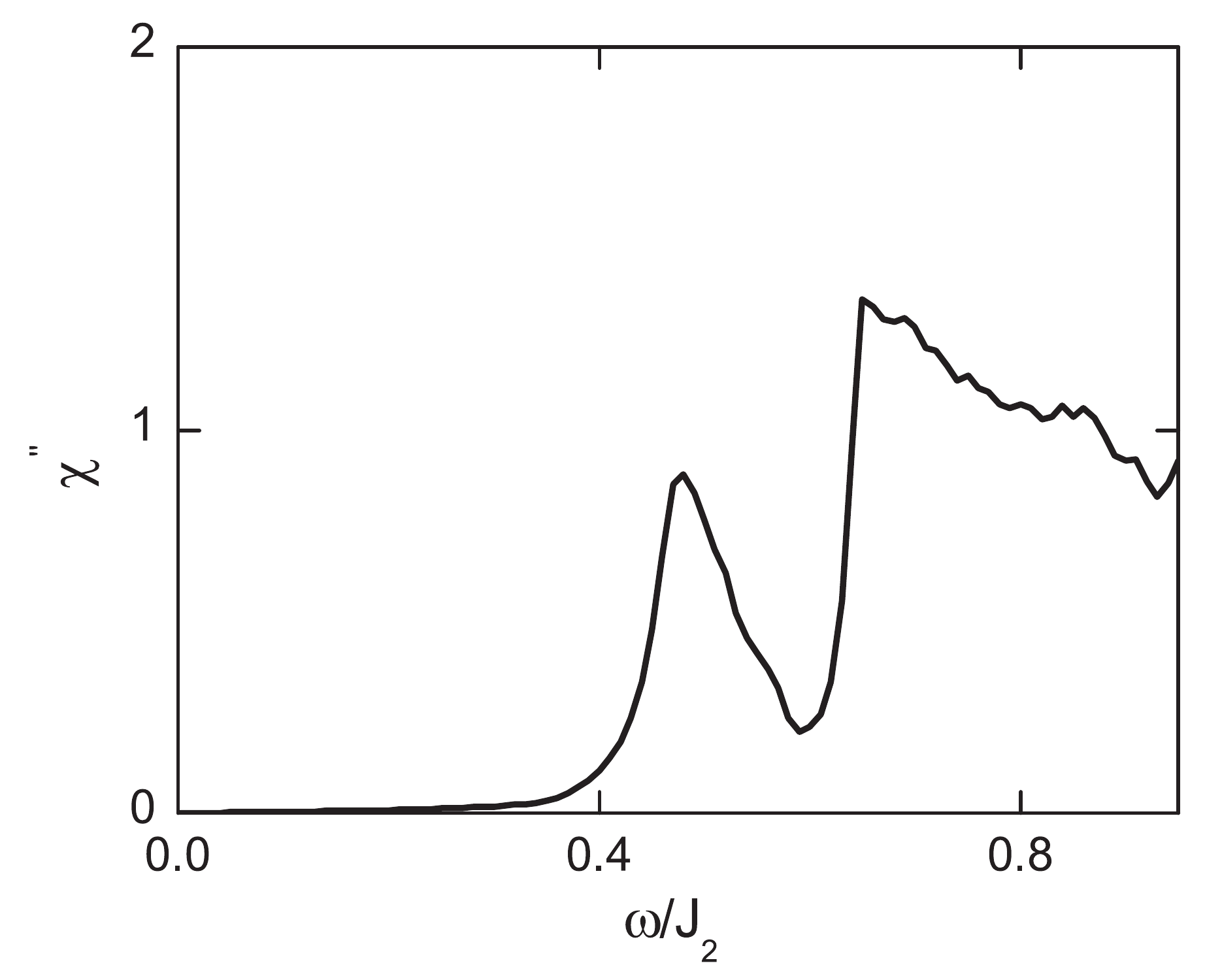}
\caption{The imaginary part of the dynamical susceptibility, $\chi^{\prime\prime}(\mathbf{Q},\omega)$ at $\mathbf{Q}=(\pi,0)$ in the SC  phase
obtained from a five-orbital t-$J_1$-$J_2$ model. 
For the case of sufficiently large gap anisotropy shown here, two resonance peaks are obtained.
}\label{Fig:5} \end{figure}

The above picture \cite{RYu_theory}
is supported by our theoretical calculation of the
dynamical spin susceptibility in the SC state of a multiorbital
$t-J_1-J_2$ model~\cite{Goswami,Yu11}. The Hamiltonian reads $H=H_0 + H_{\mathrm{int}}$. 
Here, $H_0$ contains a five-orbital tight-binding model adapted from Ref.~\cite{Graser10}. 
We have modified some tight-binding parameters such that the bandstructure better fits 
to the density function theory results on NaFeAs. The interaction part $H_{\mathrm{int}}$ includes matrix 
$J_1-J_2$ couplings.
Figure~\ref{Fig:5} shows the calculated imaginary part of the susceptibility, $\chi^{\prime\prime}(\mathbf{Q},\omega)$.
Indeed we find two resonance peaks when the gap anisotropy is large,  
which 
turn 
into one sharp peak when the gap anisotropy 
is reduced.

We now turn to several remarks. First, in the underdoped regime where the SC and AF states coexist, 
a reconstruction of Fermi surface in the AF state may in principle cause a
SC gap anisotropy.
However, 
this mechanism is unlikely
because ARPES observes neither the Fermi surface reconstruction 
 nor any gap anisotropy on the hole Fermi pocket
 in the underdoped 
NaFe$_{1-x}$Co$_x$As~\cite{qqge}.
Second, one may in principle consider the double spin resonances as originating from the
quasiparticle excitations between two different hole and electron Fermi pockets with different SC gaps.
However, such an effect would lead to spin resonances at
different wave vectors due to mismatched Fermi surfaces \cite{tdas11}.
This is unlike our observation here that
both resonances appear at the same commensurate wave vector.
Third, we have emphasized the orbital selectivity in understanding the data. Through a spin-orbit coupling,
this orbital-dependent effect may also lead to a spin anisotropy in the fluctuation spectrum.

In conclusion, we use INS to find two resonances at the same commensurate 
AF wave vector for the underdoped NaFe$_{0.985}$Co$_{0.015}$As,
 but only one resonance for the overdoped SC NaFe$_{0.955}$Co$_{0.045}$As. 
This is different from the $c$-axis dispersion of the resonance in electron-doped
BaFe$_{1.9}$Ni$_{0.1}$As$_2$ \cite{schi09} and hole-doped 
copper oxide superconductor YBa$_2$Cu$_3$O$_{6.85}$ \cite{pailhes}. 
 The doping evolution 
 of the spin resonance coincides with that of the SC gap anisotropy in ARPES experiments. 
 Our experimental discoveries, together with our theoretical analysis, suggest that both properties 
arise from the orbital dependence of the SC pairing. This provides evidence that the orbital selectivity
plays an important role in understanding the SC pairing of the multiorbital electrons in the iron pnictides.
Because the multiplicity of electron orbitals is a distinct feature of the iron-based superconductors
and likely makes a major contribution to their superconducting pairing, our results will be important 
to the eventual understanding of superconductivity in these and related materials.

We thank C. Redding and Scott Carr for their help in sample making process. 
The work at Rice/UTK was supported by the US DOE, BES, through contract DE-FG02-05ER46202 (P.D.).
Work at Rice University was supported by the NSF Grant No. DMR-1006985 
and the Robert A. Welch Foundation Grant No. C-1411 (Q.S.).
C.L.Z and T.E are partially supported by the US DOE BES through the EPSCoR grant, DE-FG02-08ER46528.
The work at the High Flux Isotope Reactor was partially supported by the Division of Scientific User Facilities, US DOE, BES.

\end{document}